\documentclass[manuscript,screen]{acmart}

\setcopyright{acmlicensed}
\copyrightyear{2025}
\acmYear{2025}
\acmConference[CHI '25]{ACM CHI Conference on Human Factors in Computing Systems}{May 2025}{Yokohama, Japan}

\begin{document}

\title{Clones in the Machine: A Feminist Critique of Agency in Digital Cloning}

\author{Siân Brooke}
\email{s.j.m.brooke@uva.nl}
\affiliation{
  \institution{Digital Interactions Lab, Informatics Institute, University of Amsterdam}
  \city{Amsterdam}
  \country{Netherlands}
}

\begin{abstract}
This paper critiques digital cloning in academic research, highlighting how it exemplifies AI solutionism. Digital clones, which replicate user data to simulate behavior, are often seen as scalable tools for behavioral insights. However, this framing obscures ethical concerns around consent, agency, and representation. Drawing on feminist theories of agency, the paper argues that digital cloning oversimplifies human complexity and risks perpetuating systemic biases. To address these issues, it proposes decentralized data repositories and dynamic consent models, promoting ethical, context-aware AI practices that challenge the reductionist logic of AI solutionism.
\end{abstract}


\maketitle
\section{Introduction}
In September 2024, a viral post \cite{ewe_goodbye_2024} circulated on Meta platforms, claiming that reposting a disclaimer would prevent Meta from using user data to train large language models (LLMs):

\begin{quote}
\textit{“Goodbye Meta AI, […] I do not give Meta or anyone else permission to use any of my personal data, profile information, or photos.”}
\end{quote}

Despite its popularity, Meta confirmed the disclaimer had no legal effect \cite{ewe_goodbye_2024}, exposing the gap between perceived and actual user agency in artificial intelligence (AI) systems. Digital replicas of users captured in their data persist without consent, raising questions about data ownership and autonomy. Similar tensions surfaced in 2023, when users on Stack Overflow and Reddit \footnote{https://www.techzine.eu/blogs/applications/120084/openai-deal-with-reddit-shows-complete-lack-of-user-agency-over-content/} staged data deletion protests after announcements that their contributions would be used for LLM training. These protests represent refusals to participate in systems that commodify digital labor, highlighting demands for digital autonomy.

At the center of these debates are digital clones—computational agents that replicate users’ digital histories (e.g., posts, interactions) to simulate and predict behaviors \cite{puri_digital_2024}. Touted as the “Holy Grail” for understanding human behavior\cite{truby_human_2021}, digital clones reflect a broader trend of AI solutionism, which assumes technical solutions can resolve complex social issues. For example, Schmidt et al. \cite{schmidt_simulating_2024} propose how LLMs can generate personas that mimic human characteristics and respond to surveys in Human-Computer Interaction (HCI) research, promising scalable insights without involving real participants.

However, this framing obscures significant ethical concerns. Digital clones do more than aggregate data—they simulate behaviors and agency, risking the reduction of complex human identities to predictive models. Assumptions of representational accuracy overlook relational and contextual dynamics that shape behavior. These issues are particularly pressing when viewed through feminist theories of agency \cite{schwartz_gendered_2019, toupin_shaping_2024}, which emphasize that autonomy is relational and shaped by power structures. This position paper examines these ethical challenges, focusing specifically on researcher-driven digital cloning in simulation studies and exploring how feminist perspectives can guide more ethical, context-aware practices in AI research.

\section{Digital Clones in Research}
Digital clones replicate user-generated data to simulate behaviors and interactions. These clones emerge through two primary approaches: (1) \textbf{User-Driven Cloning}: Users voluntarily contribute data for research, providing explicit consent. (2) \textbf{Researcher-Driven Cloning}: Researchers scrape and replicate user data, typically from public platforms, without explicit consent.

I focus on researcher-driven digital clones used in simulation studies, where researchers replicate users’ digital histories to model social network dynamics, such as misinformation spread and the formation of political echo chambers. For instance, Puri et al. \cite{puri_digital_2024} developed a simulation framework combining agent-based modeling and natural language processing (NLP) to clone a misinformation-sharing network. The study downloaded histories of over 10,000 users, reconstructed the network, and manipulated algorithms to observe interaction shifts.

While these simulations offer valuable insights into polarization and information flow, they raise significant ethical concerns. Although Puri et al. \cite{puri_digital_2024} obtained ethical approval, the ethics board deemed explicit consent unnecessary due to the data’s public availability. However, this overlooks the critical distinction between publicly available data and a digital replica. Unlike passive data collection, digital clones actively simulate user agency, predicting behaviors and interactions. Assuming these clones represent human complexity overlooks relational factors shaping behavior, such as exclusion or disengagement based on gender or ethnic identity. Additionally, treating public data as freely usable without consent undermines privacy and autonomy \cite{fiesler_lawful_2020}, echoing exploitative practices like Cambridge Analytica.

\section{Feminist Critique of Digital Cloning}
Feminist academics such as Toupin \cite{toupin_shaping_2024}, Suchman \cite{wallach_agencies_2020}, and Nunes \cite{nunes_ai_2024} highlight concerns about agency loss in AI deployment. Particularly compelling is Suchman’s \cite{wallach_agencies_2020} critique of binary oppositions, such as subject/object and human/machine, which often privilege one element over another. These dichotomies obscure the relational nature of agency \cite{wallach_agencies_2020}, shaping how digital clones are understood. Rather than being passive representations, the data generated by digital clones functions as an extension of the agency created by users. This perspective underscores the need for researchers to recognize the interconnectedness of humans and machines, as agency does not reside in discrete entities but instead emerges from sociomaterial arrangements. Digital clones, therefore, must be seen as entangled within broader configurations of algorithms, datasets, and (biased) human decisions.

\section{Ethical Concerns in Digital Cloning}
Emerging feminist critiques stated above highlight ethical concerns in digital cloning, particularly around agency, representation, and power. Key assumptions in digital cloning research include: (1) Clones sufficiently capture individual complexity; (2) Anonymized data removes the need for consent. However, these assumptions obscure the relational and contextual nature of user identities. 

Drawing on Suchman’s \cite{wallach_agencies_2020} argument, it becomes clear that digital clones risk flattening nuanced interactions. For example, personal relational choices—such as expressing political opinions differently in private groups versus public threads—can shape social dynamics. Detached from these contexts, digital clones may misrepresent behaviors, treating all interactions as uniform. Similarly, clones trained on past engagement patterns may overemphasize frequently repeated behaviors while overlooking passive or evolving user interactions, reinforcing echo chambers and distorting online discourse.

Furthermore, as Toupin \cite{toupin_shaping_2024} emphasizes, digital footprints are extensions of the self, especially for marginalized communities. Replicating these selves without explicit consent challenges user agency and risks reinforcing inequalities. Persistent clones, even after data deletion requests, deny users the right to be forgotten.

\textbf{Reinforcing Inequalities through Cloning:}  
AI-generated deepfakes show how identity replication can exploit marginalized groups \cite{karnouskos_artificial_2020}. Similarly, digital clones can misrepresent or erase diverse voices, reproducing systemic biases. The historical case of Henrietta Lacks reminds us that data extraction without consent—biological or digital—disproportionately harms oppressed groups. Feminist perspectives demand recognizing that consent is a continuous negotiation shaped by power dynamics.

\textbf{Behavioral Manipulation and Profiling:}  
Nunes’ \cite{nunes_ai_2024} posthuman feminist framework highlights how AI systems shape user behavior. Digital clones can enable behavioral manipulation through algorithmic nudges, normalizing subtle control forms. Predictive profiling risks reinforcing biases, marginalizing vulnerable populations. When clones persist despite deletion requests, users lose control over their representations, raising questions about data sovereignty.

\section{Re-imagining Ethical Frameworks for Digital Cloning}
\textbf{Decentralized Data Donation Repositories:}  
To address data power imbalances, decentralized, ethically governed repositories are essential. These repositories would enable voluntary, community-driven contributions under transparent conditions, operating exclusively for non-commercial academic research. This approach aligns with feminist HCI methodologies, emphasizing participatory design and distributed agency.

\textbf{Dynamic Consent Models with Participatory Dashboards:}  
Dynamic consent models, facilitated by interactive dashboards, are essential to restoring user agency \cite{chen_designing_2024}. These systems notify users when data is used, provide summaries of outcomes, and allow consent withdrawal from future studies at any time. Such frameworks align with Nunes’ \cite{nunes_ai_2024} call for posthuman understandings of authorship, recognizing that users and AI co-create knowledge. By enabling ongoing consent negotiation, these dashboards challenge power asymmetries between researchers and participants.

\section{Conclusion}
Digital cloning in academic research offers behavioral insights but carries ethical risks when conducted without explicit consent. The assumption that anonymized data negates ethical oversight fails to recognize how digital identities remain extensions of the self \cite{toupin_shaping_2024}. Protests on platforms like Reddit and Stack Overflow reflect broader demands for autonomy in digital environments. Academic researchers must recognize that research methods are not ethically neutral. Digital cloning research should be exposed to human-participant ethical approval, acknowledging risks to user agency, representation, and consent. By adopting dynamic consent models and decentralized data repositories, researchers can pursue more ethical AI practices that respect human identity complexities.

\begin{acks}
The author thanks the Digital Interactions Lab for their support and to Dr Katja Rogers for her constructive feedback.
\end{acks}

\bibliographystyle{ACM-Reference-Format}
\bibliography{sample-base}


\begin{thebibliography}{11}


\ifx \showCODEN    \undefined \def \showCODEN     #1{\unskip}     \fi
\ifx \showISBNx    \undefined \def \showISBNx     #1{\unskip}     \fi
\ifx \showISBNxiii \undefined \def \showISBNxiii  #1{\unskip}     \fi
\ifx \showISSN     \undefined \def \showISSN      #1{\unskip}     \fi
\ifx \showLCCN     \undefined \def \showLCCN      #1{\unskip}     \fi
\ifx \shownote     \undefined \def \shownote      #1{#1}          \fi
\ifx \showarticletitle \undefined \def \showarticletitle #1{#1}   \fi
\ifx \showURL      \undefined \def \showURL       {\relax}        \fi
\providecommand\bibfield[2]{#2}
\providecommand\bibinfo[2]{#2}
\providecommand\natexlab[1]{#1}
\providecommand\showeprint[2][]{arXiv:#2}

\bibitem[Chen et~al\mbox{.}(2024)]%
        {chen_designing_2024}
\bibfield{author}{\bibinfo{person}{Yida Chen}, \bibinfo{person}{Aoyu Wu}, \bibinfo{person}{Trevor DePodesta}, \bibinfo{person}{Catherine Yeh}, \bibinfo{person}{Kenneth Li}, \bibinfo{person}{Nicholas~Castillo Marin}, \bibinfo{person}{Oam Patel}, \bibinfo{person}{Jan Riecke}, \bibinfo{person}{Shivam Raval}, \bibinfo{person}{Olivia Seow}, \bibinfo{person}{Martin Wattenberg}, {and} \bibinfo{person}{Fernanda Viégas}.} \bibinfo{year}{2024}\natexlab{}.
\newblock \bibinfo{title}{Designing a {Dashboard} for {Transparency} and {Control} of {Conversational} {AI}}.
\newblock
\href{https://doi.org/10.48550/arXiv.2406.07882}{doi:\nolinkurl{10.48550/arXiv.2406.07882}}
\newblock
\shownote{arXiv:2406.07882 [cs]}.


\bibitem[Ewe(2024)]%
        {ewe_goodbye_2024}
\bibfield{author}{\bibinfo{person}{Koh Ewe}.} \bibinfo{year}{2024}\natexlab{}.
\newblock \bibinfo{title}{‘{Goodbye} {Meta} {AI}’ {Is} a {Privacy} {Hoax}}.
\newblock
\urldef\tempurl%
\url{https://time.com/7024218/fact-check-goodbye-meta-ai-privacy-hoax-instagram-viral-copypasta/}
\showURL{%
\tempurl}


\bibitem[Fiesler(2020)]%
        {fiesler_lawful_2020}
\bibfield{author}{\bibinfo{person}{Casey Fiesler}.} \bibinfo{year}{2020}\natexlab{}.
\newblock \showarticletitle{Lawful {Users}: {Copyright} {Circumvention} and {Legal} {Constraints} on {Technology} {Use}}. In \bibinfo{booktitle}{\emph{Proceedings of the 2020 {CHI} {Conference} on {Human} {Factors} in {Computing} {Systems}}}. \bibinfo{publisher}{ACM}, \bibinfo{address}{Honolulu HI USA}, \bibinfo{pages}{1--11}.
\newblock
\showISBNx{978-1-4503-6708-0}
\href{https://doi.org/10.1145/3313831.3376745}{doi:\nolinkurl{10.1145/3313831.3376745}}


\bibitem[Karnouskos(2020)]%
        {karnouskos_artificial_2020}
\bibfield{author}{\bibinfo{person}{Stamatis Karnouskos}.} \bibinfo{year}{2020}\natexlab{}.
\newblock \showarticletitle{Artificial {Intelligence} in {Digital} {Media}: {The} {Era} of {Deepfakes}}.
\newblock \bibinfo{journal}{\emph{IEEE Transactions on Technology and Society}} \bibinfo{volume}{1}, \bibinfo{number}{3} (\bibinfo{date}{Sept.} \bibinfo{year}{2020}), \bibinfo{pages}{138--147}.
\newblock
\showISSN{2637-6415}
\href{https://doi.org/10.1109/TTS.2020.3001312}{doi:\nolinkurl{10.1109/TTS.2020.3001312}}
\newblock
\shownote{Conference Name: IEEE Transactions on Technology and Society}.


\bibitem[Nunes(2024)]%
        {nunes_ai_2024}
\bibfield{author}{\bibinfo{person}{Rafaela Nunes}.} \bibinfo{year}{2024}\natexlab{}.
\newblock \showarticletitle{{AI}, a {Tool} or an {Author}? {A} {Posthuman} {Feminist} {Perspective} on the {Agency} of {Gen}-{AI} in {Creative} {Practices}}.
\newblock \bibinfo{journal}{\emph{Augmented Human Research}} \bibinfo{volume}{9}, \bibinfo{number}{1} (\bibinfo{date}{Dec.} \bibinfo{year}{2024}), \bibinfo{pages}{8}.
\newblock
\showISSN{2365-4317, 2365-4325}
\href{https://doi.org/10.1007/s41133-024-00074-8}{doi:\nolinkurl{10.1007/s41133-024-00074-8}}


\bibitem[Puri et~al\mbox{.}(2024)]%
        {puri_digital_2024}
\bibfield{author}{\bibinfo{person}{Prateek Puri}, \bibinfo{person}{Gabriel Hassler}, \bibinfo{person}{Sai Katragadda}, {and} \bibinfo{person}{Anton Shenk}.} \bibinfo{year}{2024}\natexlab{}.
\newblock \showarticletitle{Digital cloning of online social networks for language-sensitive agent-based modeling of misinformation spread}.
\newblock \bibinfo{journal}{\emph{PLOS ONE}} \bibinfo{volume}{19}, \bibinfo{number}{6} (\bibinfo{date}{June} \bibinfo{year}{2024}), \bibinfo{pages}{e0304889}.
\newblock
\showISSN{1932-6203}
\href{https://doi.org/10.1371/journal.pone.0304889}{doi:\nolinkurl{10.1371/journal.pone.0304889}}
\newblock
\shownote{Publisher: Public Library of Science}.


\bibitem[Schmidt et~al\mbox{.}(2024)]%
        {schmidt_simulating_2024}
\bibfield{author}{\bibinfo{person}{Albrecht Schmidt}, \bibinfo{person}{Passant Elagroudy}, \bibinfo{person}{Fiona Draxler}, \bibinfo{person}{Frauke Kreuter}, {and} \bibinfo{person}{Robin Welsch}.} \bibinfo{year}{2024}\natexlab{}.
\newblock \showarticletitle{Simulating the {Human} in {HCD} with {ChatGPT}: {Redesigning} {Interaction} {Design} with {AI}}.
\newblock \bibinfo{journal}{\emph{Interactions}} \bibinfo{volume}{31}, \bibinfo{number}{1} (\bibinfo{date}{Jan.} \bibinfo{year}{2024}), \bibinfo{pages}{24--31}.
\newblock
\showISSN{1072-5520, 1558-3449}
\href{https://doi.org/10.1145/3637436}{doi:\nolinkurl{10.1145/3637436}}


\bibitem[Schwartz and Neff(2019)]%
        {schwartz_gendered_2019}
\bibfield{author}{\bibinfo{person}{Becca Schwartz} {and} \bibinfo{person}{Gina Neff}.} \bibinfo{year}{2019}\natexlab{}.
\newblock \showarticletitle{The gendered affordances of {Craigslist} “new-in-town girls wanted” ads}.
\newblock \bibinfo{journal}{\emph{New Media \& Society}} \bibinfo{volume}{21}, \bibinfo{number}{11-12} (\bibinfo{date}{Nov.} \bibinfo{year}{2019}), \bibinfo{pages}{2404--2421}.
\newblock
\showISSN{1461-4448, 1461-7315}
\href{https://doi.org/10.1177/1461444819849897}{doi:\nolinkurl{10.1177/1461444819849897}}


\bibitem[Suchman(2020)]%
        {wallach_agencies_2020}
\bibfield{author}{\bibinfo{person}{Lucy Suchman}.} \bibinfo{year}{2020}\natexlab{}.
\newblock \showarticletitle{Agencies in {Technology} {Design}: {Feminist} {Reconfigurations}*}.
\newblock In \bibinfo{booktitle}{\emph{Machine {Ethics} and {Robot} {Ethics}} (\bibinfo{edition}{1} ed.)}, \bibfield{editor}{\bibinfo{person}{Wendell Wallach} {and} \bibinfo{person}{Peter Asaro}} (Eds.). \bibinfo{publisher}{Routledge}, \bibinfo{pages}{361--375}.
\newblock
\showISBNx{978-1-00-307499-1}
\href{https://doi.org/10.4324/9781003074991-32}{doi:\nolinkurl{10.4324/9781003074991-32}}


\bibitem[Toupin(2024)]%
        {toupin_shaping_2024}
\bibfield{author}{\bibinfo{person}{Sophie Toupin}.} \bibinfo{year}{2024}\natexlab{}.
\newblock \showarticletitle{Shaping feminist artificial intelligence}.
\newblock \bibinfo{journal}{\emph{New Media \& Society}} \bibinfo{volume}{26}, \bibinfo{number}{1} (\bibinfo{date}{Jan.} \bibinfo{year}{2024}), \bibinfo{pages}{580--595}.
\newblock
\showISSN{1461-4448}
\href{https://doi.org/10.1177/14614448221150776}{doi:\nolinkurl{10.1177/14614448221150776}}
\newblock
\shownote{Publisher: SAGE Publications}.


\bibitem[Truby and Brown(2021)]%
        {truby_human_2021}
\bibfield{author}{\bibinfo{person}{Jon Truby} {and} \bibinfo{person}{Rafael Brown}.} \bibinfo{year}{2021}\natexlab{}.
\newblock \showarticletitle{Human digital thought clones: the {Holy} {Grail} of artificial intelligence for big data}.
\newblock \bibinfo{journal}{\emph{Information \& Communications Technology Law}} \bibinfo{volume}{30}, \bibinfo{number}{2} (\bibinfo{date}{May} \bibinfo{year}{2021}), \bibinfo{pages}{140--168}.
\newblock
\showISSN{1360-0834}
\href{https://doi.org/10.1080/13600834.2020.1850174}{doi:\nolinkurl{10.1080/13600834.2020.1850174}}
\newblock
\shownote{Publisher: Routledge}.


\end{thebibliography}

\end{document}